\documentclass[letterpaper, 10 pt, conference]{ieeeconf}  % Comment this line out if 4you need a4paper

\IEEEoverridecommandlockouts                              % This command is only needed if 
\usepackage{graphicx}
\usepackage{booktabs}
\usepackage{epsfig} % for postscript graphics files
\usepackage{amsmath} % assumes amsmath package installed
\usepackage{amssymb}  % assumes amsmath package installed
\usepackage{mathrsfs}
\usepackage{cleveref}
\usepackage{color}
\usepackage{calc}
\usepackage{dsfont}
\usepackage{bm}
\usepackage{float}
\newtheorem{lemma}{Lemma}
\newtheorem{theorem}{Theorem}
\newtheorem{proposition}{Proposition}
\newtheorem{problem}{Problem}

\newtheorem{corollary}{Corollary}

\newenvironment{customproof}[1][Proof]
{\par\indent\textit{#1:} }
{\hfill$\blacksquare$\par}

\usepackage{xspace}

\newcommand{\paramTrueDelay}{\ensuremath{D}\xspace}

\DeclareMathOperator*{\argmax}{arg\,max}
\DeclareMathOperator*{\argmin}{arg\,min}

\usepackage{etoolbox}
\makeatletter
\patchcmd{\@makecaption}
  {\scshape}
  {}
  {}
  {}
\makeatletter
\patchcmd{\@makecaption}
  {\\}
  {.\ }
  {}
  {}
\makeatother

\usepackage{amsfonts} 
\title{\LARGE \bf
Minimizing Conservatism in Safety-Critical Control for \\ Input-Delayed Systems via Adaptive Delay Estimation}

\author{Yitaek Kim$^{1}$, Ersin Daş$^{2}$,  Jeeseop Kim$^{2}$,  Aaron D. Ames$^{2}$, Joel W. Burdick$^{2}$ and Christoffer Sloth$^{1}$
\thanks{$^{1}$Authors are with the Maersk Mc-Kinney Moller Institute, University of Southern Denmark, Denmark {\tt\small \{yik,chsl\}@mmmi.sdu.dk}}
\thanks{$^{2}$Authors are with the Department of Mechanical and Civil Engineering, California Institute of Technology, Pasadena, CA 91125, USA,
 {\tt\small \{ersindas, jeeseop, ames, jburdick\}@caltech.edu}}
}

\usepackage{tikz}

\newcommand\submittedtext{%
  \footnotesize \textcopyright \text{ }2025 IEEE.  Personal use of this material is permitted.  Permission from IEEE must be obtained for all other uses, in any current or future media, including reprinting/republishing this material for advertising or promotional purposes, creating new collective works, for resale or redistribution to servers or lists, or reuse of any copyrighted component of this work in other works.}
  
\newcommand\submittednotice{%
\begin{tikzpicture}[remember picture,overlay]
\node[anchor=south,yshift=10pt] at (current page.south) {\fbox{\parbox{\dimexpr\textwidth-\fboxsep-\fboxrule\relax}{\submittedtext}}};
\end{tikzpicture}%
}

\begin{document}
\maketitle
\submittednotice
\thispagestyle{empty}
\pagestyle{empty}

%%%%%%%%%%%%%%%%%%%%%%%%%%%%%%%%%%%%%%%%%%%%%%%%%%%%%%%%%%%%%%%%%%%%%%%%%%%%%%%%
\begin{abstract}
Input delays affect systems such as wirelessly connected autonomous vehicles, and may lead to safety violations. One promising way to ensure safety in the presence of delay is to employ control barrier functions (CBFs), and extensions thereof that account for uncertainty: delay adaptive CBFs (DaCBFs). This paper proposes an online adaptive safety control framework for reducing the conservatism of DaCBFs. The main idea is to reduce the maximum delay estimation error bound so that the state prediction error bound is monotonically non-increasing. To this end, we first leverage both delay estimation and the estimation error bound of a disturbance observer to derive an upper bound on the current state prediction error from the previous state. Second, we design two nonlinear programs to update the maximum delay estimation error bound satisfying the obtained state prediction error bound previously and afterwards update the maximum error bound of the future state prediction used in DaCBFs. The proposed method ensures the maximum state prediction error bound with the delay estimation is monotonically non-increasing, yielding less conservatism in DaCBFs. We verify the proposed method in an automated connected truck application, showing that the proposed method reduces the conservatism of DaCBFs.
\end{abstract}
%Input delays affect systems such as teleoperation and wirelessly autonomous connected vehicles, and may lead to safety violations. One promising way to ensure safety in the presence of delay is to employ control barrier functions (CBFs), and extensions thereof that account for uncertainty: delay adaptive CBFs (DaCBFs). This paper proposes an online adaptive safety control framework for reducing the conservatism of DaCBFs. The main idea is to reduce the maximum delay estimation error bound so that the state prediction error bound is monotonically non-increasing. To this end, we first leverage the estimation error bound of a disturbance observer to bound the state prediction error. Second, we design two nonlinear programs to update the maximum delay estimation error bound satisfying the state prediction error bound, and subsequently update the maximum state prediction error bound used in DaCBFs. The proposed method ensures the maximum state prediction error bound is monotonically non-increasing, yielding less conservatism in DaCBFs. We verify the proposed method in an automated connected truck application, showing that the proposed method reduces the conservatism of DaCBFs.

%%%%%%%%%%%%%%%%%%%%%%%%%%%%%%%%%%%%%%%%%%%%%%%%%%%%%%%%%%%%%%%%%%%%%%%%%%%%%%%%
\section{Introduction}\label{sec:introduction}
As the demand for safety guarantees has increased in real-world control systems, safety-critical control design for systems with uncertain dynamics has been widely studied across many practical applications. Control barrier functions (CBFs) \cite{Ames2017CBFQP} are a promising method to accomplish safety, and they have recently been extended to robust \cite{lopez2020robust}, data-driven \cite{YK2024GPRaCBF} approaches in the context of delay-free systems, showing high scalability and computational efficiency when realized as a safety-filter \cite{ames2019control} framed as a quadratic program (QP). However, time delays can often happen in urban application settings such as a connected automated car \cite{Tamas2022ISSwithInputDelay}, teleoperation-based controls \cite{Emmanuel2009}, leading to safety violations; thus, time delays must be incorporated into the CBFs-based control design.

To address the impact of delays on safety, recent works extend CBFs to input-delayed linear \cite{Jankovic2018} and nonlinear systems \cite{Singletary2020, Molnar2022} using state prediction-based feedback controller \cite{Karafyllis2017}. For more complex cases, the robust safety-critical control scheme that considers known input delay and additional disturbances is proposed to ensure safety guarantees, and the scheme is efficiently demonstrated with a connected automated vehicle control system example \cite{Tamas2022ISSwithInputDelay}. 
However, these works assumed that the input delay was already known. For unknown input delay, integral quadratic constraints techniques to bound the disturbances caused by the delay are integrated with a CBFs-based controller \cite{Seiler2022}, which is extended to a tube-based CBF to establish robustness with respect to the unknown input delays \cite{Quan2023}. The recent work, delay adaptive CBFs (DaCBFs) \cite{YK2024ECC} combine delay estimation and CBFs to achieve robustness against worst-case uncertainties in state prediction and unknown input delay. Nevertheless, the proposed controller is too conservative.

One promising way to decrease conservatism is to use disturbance observers. Several works have recently proposed to estimate external disturbances and then compensate for them directly. An estimation error quantified observer, the general version of state observers, is combined with a function approximation \cite{Huang2001FAT} and CBFs to ensure safety in the presence of the state uncertainty \cite{Wang2022ACC_EEQ}. A high-gain disturbance observer is also designed to estimate the impacts of Lie derivatives on unknown disturbances with an exponentially decreasing error bound, which is used to achieve safety in a CBF-based controller \cite{Ersin2022}.  An extension of work \cite{Ersin2022} proposes a stricter error bound condition that includes transient errors by considering the initial condition of the observer, which results in less conservatism \cite{AlanDOBCBF2023}.  In \cite{Wang2023}, a nonlinear disturbance observer proposed by \cite{Chen2004DOB} is integrated with a robust CBF, which reduces the conservatism. However, disturbance observers have not yet been effectively applied to systems with unknown input delays. 

% One promising way to decrease conservatism is to use disturbance observers. Several works have recently proposed to estimate external disturbances and then compensate for them directly. An estimation error quantified observer, the general version of traditional state observers, is combined with a function approximation technique \cite{Huang2001FAT} and CBFs to ensure safety in the presence of the state uncertainty \cite{Wang2022ACC_EEQ}. A high-gain disturbance observer is also designed to estimate the impacts of Lie derivatives on unknown disturbances with an exponentially decreasing error bound, which is used to achieve safety in a CBF-based controller \cite{Ersin2022}. An extension of work \cite{Ersin2022} proposes a stricter error bound condition that includes transient errors by considering the initial condition of the observer, which results in less conservatism \cite{AlanDOBCBF2023}. 
% In \cite{Wang2023}, a nonlinear disturbance observer proposed by \cite{Chen2004DOB} is integrated with a robust CBF, which reduces the conservatism. While disturbance observers have been combined with various safe controllers to decrease inherent conservatism, they have not yet been effectively applied to systems with unknown input delays. 
\begin{figure*}[t]
    \centering
    \includegraphics[width=1\linewidth]{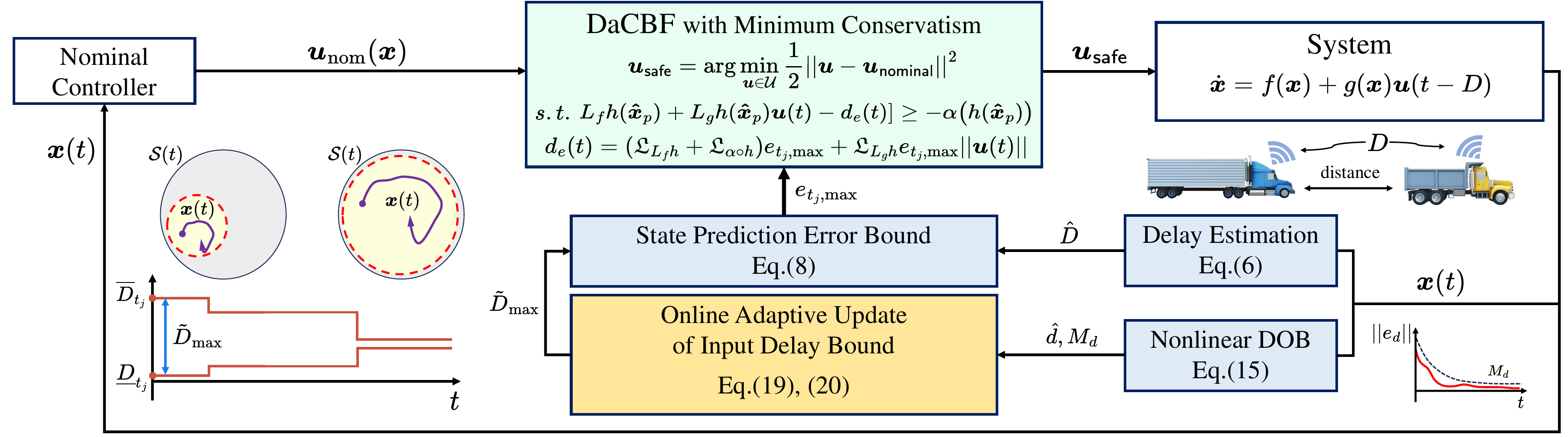}
    \caption{A block diagram of the proposed method for ensuring safety in unknown input-delayed systems. The proposed method consists of three components: delay and disturbance estimations, online adaptive update of input delay bound, and DaCBFs with minimum conservatism. The estimated input delay and disturbance are leveraged to update the bound set of input delay and the maximum state prediction error bound, $e_{t_j, \textnormal{max}}$. Given the updated maximum delay estimation error bound, $\tilde{D}_{\textnormal{max}}$ from the online adaptive algorithm \eqref{pm:nlp1} and \eqref{pm:nlp2}, $e_{t_j, \textnormal{max}}$ is updated and then used in the robust condition in DaCBFs, monotonically non-increasing over time if $\bm{u}(t)$ is not constant. The proposed method ensures less conservative than the previous work, DaCBFs \cite{YK2024ECC}.}
    \vspace{-0.4cm}
    \label{pro:framework}
\end{figure*}

In this paper, we propose an online adaptive safety-critical control framework with minimum conservatism for the unknown input-delayed system. To this end, we extend our previous framework, DaCBFs \cite{YK2024ECC}, while ensuring gradually non-increasing maximum state prediction error by using delay estimation and a disturbance observer, thereby reducing conservatism. We demonstrate that the proposed method adapts to the unknown input delay and decreases conservatism online, compared to DaCBFs in a connected automated vehicle simulation. The entire system architecture is shown in Fig.~\ref{pro:framework}. 

This paper is organized as follows. Section~\ref{sec:problem_formulation} presents the problem solved in the paper, and Section~\ref{sec:preliminary_knowledges} revisits input
delay estimation, delay adaptive control barrier functions,
and a disturbance observer as preliminaries. Section~\ref{sec:method} describes the proposed method and provides analysis about less conservatism. The proposed method is verified on the connected automated vehicles simulation in Section~\ref{sec:simulation}. Lastly, Section~\ref{sec:conclusions} concludes the paper with future works. 

\section{Problem Formulation}\label{sec:problem_formulation}
Consider the following control affine nonlinear system with a constant input delay, denoted as $D$:
\begin{equation}
\dot{\bm{x}} = f(\bm{x}) + g(\bm{x})\bm{u}(t-\paramTrueDelay), \label{pf:sys_real}
\end{equation}
where ${\bm{x} \!\in\! \mathcal{X} \!\subset\! \mathbb{R}^n}$ is the state of the system, ${f: \mathcal{X} \rightarrow \mathbb{R}^n}$ and ${g: \mathcal{X} \rightarrow \mathbb{R}^{n\times m}}$ are locally Lipschitz continuous functions, ${\bm{u} \in \mathcal{U} \subset \mathbb{R}^m}$ is the control input, and ${0\leq \underline{D} \leq D \leq \overline{D}}$. The aim of this paper is to provide safety guarantees for \eqref{pf:sys_real}, i.e. ensure that some set ${\mathcal{S}\subseteq \mathcal{X}}$ is positively invariant. At the same time, the conservatism of  DaCBFs \cite{YK2024ECC} is reduced by estimating the input delay and bounding the delay estimation error. Specifically, we address the following problem:
\begin{problem}\label{prob:pb1}
Design an online adaptive algorithm to decrease the conservatism of DaCBFs by leveraging delay estimation.
\label{pf:problem1}
\end{problem}

% Consider the following control affine nonlinear system with a constant input delay, denoted as $D$:
% \begin{equation}
% \dot{\bm{x}} = f(\bm{x}) + g(\bm{x})\bm{u}(t-\paramTrueDelay), \label{pf:sys_real}
% \end{equation}
% where ${\bm{x} \!\in\! \mathcal{X} \!\subset\! \mathbb{R}^n}$ is the state of the system, ${f: \mathcal{X} \rightarrow \mathbb{R}^n}$ and ${g: \mathcal{X} \rightarrow \mathbb{R}^{n\times m}}$ are locally Lipschitz continuous functions, ${\bm{u} \in \mathcal{U} \subset \mathbb{R}^m}$ is the control input, and ${0\leq \underline{D} \leq D \leq \overline{D}}$. The aim of this paper is to provide safety guarantees for \eqref{pf:sys_real}, i.e. ensure that some set ${\mathcal{S}\subseteq \mathcal{X}}$ is positively invariant. The work is similar to DaCBFs \cite{YK2024ECC}; however, in this work, the conservatism of the safety-enforcing control is reduced by estimating the input delay and bounding the delay estimation error. Specifically, we address the following problem:

\section{Preliminaries}\label{sec:preliminary_knowledges}
This section presents background on input delay estimation, delay adaptive control barrier functions (DaCBFs), and disturbance estimation.
\subsection{Input Delay Estimator}
\label{sec:IDE}
Estimation of input delays can be accomplished with a gradient descent algorithm similar to \cite{bresch2012}, \cite{BertinoDelay2022} based on the state prediction error. A state prediction at time $t$ is computed by forward integrating of \eqref{pf:sys_real} with the initial condition given by $\bm{x}(t-\beta)$, where ${\beta > 0}$ is a design parameter, using an estimated input delay, $\hat{D}$ \cite{BertinoDelay2022}:
\begin{align}
    &\bm{\hat{x}}_p(\delta,t,\hat{D}) \nonumber\\
    &= \bm{x}(t-\beta)+\beta \int_{0}^{\delta} {f_0(\bm{\hat{x}}_p(y,t,\hat{D}),\bm{u}_p(y,t,\hat{D}))} dy, \label{pre:state_predictor_xp}
\end{align}
where ${ \delta \in [0,1]}$ is a distribution variable, and $f_0(\cdot,\cdot)$ takes form,
\begin{align*}
&f_0(\bm{\hat{x}}_p(\delta,t,\hat{D}),\bm{u}_p(\delta,t,\hat{D})) \\ &= f(\bm{\hat{x}}_p(\delta,t,\hat{D})) + g(\bm{\hat{x}}_p(\delta,t,\hat{D}))\bm{u}_p(\delta,t,\hat{D}),
\end{align*}
%$$ f_0(\bm{\hat{x}}_p(\delta,t,\hat{D}),\bm{u}_p(\delta,t,\hat{D})) = f(\bm{\hat{x}}_p(\delta,t,\hat{D})) + g(\bm{\hat{x}}_p(\delta,t,\hat{D}))\bm{u}_p(\delta,t,\hat{D}),$$ 
and the distributed input, $\bm{u}_p(\delta,t,\hat{D})=\bm{u}(t-\hat{D}+\beta(\delta-1))$. From \eqref{pre:state_predictor_xp}, it is seen that:
\begin{align}
    & \bm{\hat{x}}_p(\delta,t,D) = \bm{x}(t-\beta(\delta-1))
    \nonumber, \\
    & \bm{\hat{x}}_p(1,t,D) = \bm{x}(t). \label{pre:true_prediction}
\end{align}
We utilize property \eqref{pre:true_prediction} to formulate the cost function \cite{BertinoDelay2022}: 
\begin{equation}
   J (t,\hat{D}) \triangleq {\frac{1}{2}\Big\vert\Big\vert \bm{\hat{x}}_p(1,t,\hat{D})-\bm{x}(t)\Big\vert\Big\vert^2}, \label{pre:cost_function}
\end{equation}
where ${\hat{D} = D}$ is a minimizer of $J$.
To obtain the adaptation law of the delay estimator, the steepest descent method \cite{gradientPolak} is used for \eqref{pre:cost_function} with gradient:
\begin{equation}
    \frac{\partial J}{\partial \hat{D}}(t,\hat{D})=\Big(\bm{\hat{x}}_p(1,t,\hat{D})-\bm{x}(t)\Big)^{\top}\frac{\partial \bm{\hat{x}}_p}{\partial \hat{D}}(1,t,\hat{D}). \label{pre:gradient_j}
\end{equation}
From \eqref{pre:gradient_j}, the input delay estimator is proposed by \cite{BertinoDelay2022} as 
\begin{equation}
    \dot{\hat{D}}(t) = \gamma \text{Proj}_{[\underline{D},\overline{D}]}\Big\{\hat{D}(t), \rho_{D}(t) \Big\},     \label{pre:update_law}
\end{equation}
% \begin{equation*}
%     \rho_{D}(t) = \frac{-\frac{\partial J}{\partial \hat{D}}(t,\hat{D})}{1+\Big\vert\Big\vert \frac{\partial \bm{\hat{x}}_p}{\partial \hat{D}}(1,t,\hat{D})\Big\vert\Big\vert^2},
% \end{equation*}
where $\rho_{D}(t) = \frac{-\frac{\partial J}{\partial \hat{D}}(t,\hat{D})}{1+\Big\vert\Big\vert \frac{\partial \bm{\hat{x}}_p}{\partial \hat{D}}(1,t,\hat{D})\Big\vert\Big\vert^2}$ and ${\gamma \!>\! 0}$ regulates the adaptation rate of the estimator, and $\text{Proj}_{[a,b]}$ ensures that the thresholds of the estimated input delay \cite{BertinoDelay2022} as
\begin{equation*}
    \text{Proj}_{[a,b]}(f,g)=g \begin{cases}
0, ~ \text{if } f = a \text{ and } g < 0 \\
0, ~ \text{if } f = b \text{ and } g > 0
\\
1, ~ \text{ otherwise.}
\end{cases}
\end{equation*}
The following lemma provides a bound on $\rho_{D}(t)$: 

% \ed{Is it possible to apply an adaptive step size method to $\gamma$? The optimization/adaptation might converge faster.}

% \ed{Is there an upper/lower bound for $\beta$? I believe this parameter indicates the level of accuracy in the system model, specifically regarding the uncertainty bound. For our problem setup, larger values of $\beta$ might be better since the only uncertainties are the delay and the state errors.}

% \ed{Can we cite \cite{bresch2012}, which proposes the delay estimator approach first, and then it was adapted by \cite{BertinoDelay2022} (The first equation in Section 9.1)?}

\begin{lemma}[\cite{BertinoDelay2022}]
\textit{If $\bm{x}(t), \bm{u}(t), $ and $\bm{\dot{u}}(t)$ are uniformly bounded, and the initial estimation error ${\tilde{D}(0) = D-\hat{D}(0)}$, where ${\tilde{D}(t) \triangleq D-\hat{D}(t)}$, is bounded by ${|\tilde{D}(0)| < \tilde{D}_{\textnormal{max}}}$, then there exists a parameter ${H\in\mathbb{R}_{+}}$ such that
    \begin{align*}
        |\rho_D(t)| \leq H,
    \end{align*}
and the estimation error $\tilde{D}(t)$ satisfies: 
    \begin{align*}
        \tilde{D}(t)\rho_D(t) \geq 0,  %\label{pre:non_increasing_cond} 
    \end{align*}
which ensures that the delay estimation error is monotonically non-increasing.} For brevity, we denote $\hat{D}(t)$ by $\hat{D}$ in the following.
\label{pre:bound_d_e}
\end{lemma}

\subsection{Delay adaptive CBF}
The state prediction \eqref{pre:state_predictor_xp} and the input delay estimator \eqref{pre:update_law} can be efficiently integrated with CBFs to ensure robust safety guarantees. To achieve this, the state prediction error between the systems with input delay $D$ and $\hat{D}$ respectively can be bounded with the help of the following theorem:  
\textit{\begin{theorem}[\cite{YK2024ECC}]
Let ${\bm{x}:\mathds{R}_+\rightarrow \mathds{R}^n}$ and  ${\bm{y}:\mathds{R}_+\rightarrow \mathds{R}^n}$ be the solutions of \eqref{pf:sys_real} with initial conditions ${\bm{x}(0) = \bm{y}(0)}$ and input delay $D$ and $\hat{D}$, respectively. Let $\mathfrak{L}_f$ and $\mathfrak{L}_g$ be Lipschitz constants of $f$ and $g$, respectively, and let the input be bounded by $||\bm{u}(t)|| \leq u_{\text{max}}$, and input difference be bounded, ${||\bm{u}(t-D)-\bm{u}(t-\hat{D})||\leq \epsilon_{\text{max}}, \quad \epsilon_{\text{max}}\in\mathbb{R}_+}$ for all $t\geq 0$. Then,
the error ${\bm{e}(t)=\bm{x}(t)-\bm{y}(t)}$ on the interval ${t\in[t_1, t_2]}$, ${0\leq t_1 < t_2}$, is also bounded as
\begin{equation}
    \vert\vert \bm{e}(t) \vert\vert \leq \epsilon_{\text{max}}\int_{t_1}^{t_2}{}e^{a(t-\tau)}\vert\vert g(\bm{y}(\tau))\vert\vert d\tau \triangleq e_{\text{max}}(t)
    \label{eqn:bound_e_dmax}
\end{equation}
where ${a \triangleq \mathfrak{L}_f +\mathfrak{L}_g(u_{\text{max}} + \epsilon_{\text{max}})}$.
\label{pre:state_pre_e_bd}
\end{theorem}}

The delay estimation error, $\tilde{D}$ is considered in the state prediction error with delay estimation; thus the following corollary is derived:
\textit{\begin{corollary}[\cite{YK2024ECC}]\label{cor:e_bound}
   Let ${||D-\hat{D}||\leq \tilde{D}_{\text{max}}}$. Then, the state prediction error ${\bm{x}(t+D)-\bm{y}(t+\hat{D})}$ is bounded as
\begin{equation}
    ||\bm{x}(t+D)-\bm{y}(t+\hat{D})|| \leq e_{\text{max}}(t+\hat{D}+\tilde{D}_{\text{max}})+\Delta y_{\text{max}}, \label{eqn:prediction_error_bd}
\end{equation}
where ${e_{\text{max}}(t+\hat{D}+\tilde{D}_{\text{max}})}$ is given in \eqref{eqn:bound_e_dmax}, and
\begin{align}
   \Delta y_{\text{max}}= \max_{\tilde{D}\in[-\tilde{D}_{\text{max}}, \tilde{D}_{\text{max}}]}||\bm{y}(t+\hat{D}+\tilde{D})-\bm{y}(t+\hat{D})||.
   \end{align}
\label{eqn:state_error_bd_to_delay}
\end{corollary}}

Let ${\mathcal{S} \triangleq \{\bm{x} \in \mathcal{X} \text{ }\vert\text{ } h(\bm{x}) \geq 0\}}$ where ${h: \mathcal{X} \rightarrow \mathbb{R}}$ is a continuously differentiable function. Then, we can establish formal robust safety guarantees, with respect to $\mathcal{S}$, for system \eqref{pf:sys_real} with the help of the following theorem: 
%robust safe controller ensures the safety of \eqref{pf:sys_real} 
%with respect to $\mathcal{S}$.
\textit{\begin{theorem}[\cite{YK2024ECC}]
    Let ${||\bm{u}(t-D)-\bm{u}(t-\hat{D})||\leq \epsilon_{\text{max}}}$, and ${||\bm{u}(t)||\leq u_{\text{max}}}$ for all ${t\geq0}$. Then, the output function ${h: \mathcal{X}  \rightarrow \mathbb{R} }$ is a \textit{delay adaptive control barrier function} for \eqref{pf:sys_real} on $\mathcal{S}$ provided that there exists an extended class $\mathcal{K}_{\infty}$ function $\alpha$ such that for all the predicted states, ${\bm{\hat{x}}_p \triangleq \bm{x}(t+\hat{D}) \in \mathcal{S}}$: 
    \begin{align}
        \sup_{\bm{u}\in\mathcal{U}}[L_{f}h(\bm{\hat{x}}_p) + L_{g}h(\bm{\hat{x}}_p)\bm{u}(t) -d_e(t)] %\nonumber \\
        &\geq -\alpha\big(h(\bm{\hat{x}}_p)\big),
        \label{pre:dacbf_def}
    \end{align}
    where %$\bm{z} = \bm{p}(\bm{x}_p)$ from \eqref{eqn:sensor_measurement}.
    \begin{align*}
        d_e(t) \triangleq (\mathfrak{L}_{L_fh}+ \mathfrak{L}_{\alpha\circ h})e_{p,\text{max}}+\mathfrak{L}_{L_gh}e_{p,\text{max}}||\bm{u}(t)||, \label{theo1_d}
    \end{align*}
    with ${e_{p,\text{max}} \triangleq e_{\text{max}}(t+\hat{D}+\tilde{D}_{\text{max}})+\Delta y_{\text{max}}}$. Then, if $h$ is a DaCBF and there exists an input ${\bm{u}\in\mathcal{U}}$ satisfying \eqref{pre:dacbf_def}, then the system \eqref{pf:sys_real} is safe with respect to $\mathcal{S}$, for all ${ t \geq D}$.
\end{theorem}}

% \subsection{High Gain Observer}
% Consider the following dynamical system with unknown time-varying dynamics, $w_d \in \mathbb{R}$: 
% \begin{equation}
%     \dot{z}_d = v_d + w_d, \label{hgoest0}
% \end{equation}
% where $z_d \in \mathbb{R}$ and $v_d \in \mathbb{R}$ are measurable variables. To estimate $w_d$, we first define the estimator as follows \cite{STOTSKY20021371}:
% \begin{align}
%     \hat{w}_d &= k_d z_d-\varepsilon_d, \label{hgoest1}\\ 
%     \dot{\varepsilon}_d &= -k_d\varepsilon_d + k_d v_d + k^2_d z_d, \label{hgoest2}\\
%     e_d &= w_d + \varepsilon_d -k_d z_d, \label{hgoest3}
% \end{align}
% where $\hat{w}_d \in \mathbb{R}$ is the estimated unknown dynamics, and $k_d \in \mathbb{R}_{>0}$ is the observer gain, and $\varepsilon_d \in \mathbb{R}$ is an auxiliary variable, and $e_d = w_d - \hat{w}_d \in \mathbb{R}$ is the estimation error. We assume that the estimator from \eqref{hgoest0}-\eqref{hgoest3} is an \textit{estimation error quantified observer} by \cite{Wang2022ACC_EEQ} satisfying:
% \begin{equation}
%     |e_d| \leq M_d(t,w_d,\hat{w}_d), \quad \forall t \geq 0, \label{pre:est_bounds_m}
% \end{equation}
% where $M_d(t,w_d,\hat{w}_d)$ is the bound of the estimation error. And we further assume $\dot{w}_d$ is globally bounded as $|\dot{w}_d| \leq b_1$. Then, the error, $e_d$ is bounded as follows \cite{STOTSKY20021371}:
% \begin{equation}
%     |e_d| \leq \sqrt{e_d(0)^2e^{-k_d t} + \frac{b^2_1}{k^2_d}} = M_d \label{pre:bound_estimation_error}
% \end{equation}

\subsection{Disturbance Observer}
We consider a control affine nonlinear dynamical system with an unknown input disturbance, $\bm{d}(t)$ with ${\bm{d} :\mathds{R}_+\rightarrow \mathds{R}^m}$:
\begin{equation}
\dot{\bm{x}} = f(\bm{x}) + g(\bm{x})\big(\bm{u}+\bm{d}(t)\big),
\end{equation}
such that the disturbance and its derivative are bounded by some known constants ${w_0, w_1 > 0}$ as ${||\bm{d}(t)|| \leq w_0}$, ${||\dot{\bm{d}}(t)|| \leq w_1}$.
To estimate $\bm{d}$, \cite{DOBChens2004} proposes a nonlinear disturbance observer with the following structure:
\begin{align}
    \hat{\bm{d}} &= \bm{z} + \alpha_h P(\bm{x}),  \label{pre:est_dob} \\
    \dot{\bm{z}} &= -\alpha_h L_d(x) \big(f(\bm{x}) + g(\bm{x})(\bm{u} + \hat{\bm{d}} )\big), \nonumber 
\end{align}
where ${\hat{\bm{d}} \in \mathds{R}^m}$, is the estimated disturbance, ${\bm{z} \in \mathds{R}^m}$ is an auxiliary variable, $\alpha_h$ is a non-negative tuning parameter, and $P(\bm{x})$ is an estimation gain function satisfying $\frac{\partial P(\bm{x})}{\partial \bm{x}} = L_d(\bm{x})$, and $L_d(\bm{x})$ is the designed gain function \cite{Mohammadi} satisfying: ${\bm{e}_d^{\top}\bm{e}_d \leq \bm{e}_d^{\top}L_d(\bm{x})g(\bm{x})\bm{e}_d}$, where ${\bm{e}_d \triangleq \bm{d} -\hat{\bm{d}}}$ is the disturbance estimation error. Then, the disturbance estimation error is uniformly bounded as \cite{Wang2023}:
\begin{equation}
    ||\bm{e}_d(t)|| \!\leq\! \sqrt{\frac{2ck||e_d(0)||^2e^{-2kt}+w^2_1(1-e^{-2kt})}{2ck}}\triangleq M_d(t),  \label{pre:est_bounds_m}
\end{equation}
where ${k \triangleq \alpha_h - \frac{c}{2}}$, with ${0 < c <2\alpha_h }$. %is a parameter. % satisfying $c<2\alpha$.

%errro$\bm{e}_d$ of the disturbance observer in \eqref{pre:est_dob} 

%In [26], the condition is stated as ${\bm{e}_d \triangleq \bm{d} - \alpha_h \hat{\bm{d}}}$, as shown in their Equation 8. While assuming $\alpha_h = 1$ may be acceptable, it imposes more restrictions when defining \( L \).
\section{Main Results}\label{sec:method}
In this section, we propose an online adaptive algorithm to update the maximum bound of the delay estimation error; this makes it possible to reduce the conservatism of DaCBFs.

\subsection{Disturbance Observer with Input Delay}
We show how a disturbance observer can be used in the time-delayed system \eqref{pf:sys_real} in order to estimate the disturbance caused by the unknown input delay. Consider the input-delayed system \eqref{pf:sys_real}, which is equivalent to the following system:
\begin{equation}
\label{eq:sys_unc}
\dot{\bm{x}} = f(\bm{x}) + g(\bm{x})\bm{u}(t-\hat{D}) + g(\bm{x})\bm{d}(t)
\end{equation}
where $\bm{d}(t) = \bm{u}(t-D) - \bm{u}(t-\hat{D})$. 
Inspired by \eqref{pre:est_dob}, we propose the following disturbance observer:
\begin{align}
    \hat{\bm{d}} &= \bm{z} + \alpha_h P(\bm{x})  \label{pm:dob_delay} \\
    \dot{\bm{z}} &= -\alpha_h L_d\Big(f(\bm{x}) + g(\bm{x})\big(\bm{u}(t-\hat{D}) + \hat{\bm{d}}\big)\Big). \nonumber 
\end{align}
We use the estimated disturbance, $\hat{\bm{d}}$, and the estimation error upper bound to define a constraint to update the maximum and minimum bounds of $D$ in the following.

\subsection{Online Adaptive Update of Input Delay Bounds}
To update the maximum delay estimation error bound, we first attempt to find maximum and minimum bounds on the input delay in an online fashion and define a sequence of sets as
 \begin{equation}
      \Xi_{t_j} \triangleq \Big\{D \in \Xi_{{t_j-1}}\text{ }\Big|\text{ } \underline{D}^{t_j} \leq D \leq \overline{D}^{t_j} \Big\}, \label{pm:max_bounded_set}
 \end{equation}
  with ${t_j\in\mathbb{N}}$ and ${\Xi_0 = [\underline{D}^{0}, \overline{D}^{0}]}$. When given the set, $\Xi_{t_j}$, we are able to update the maximum bound of the delay estimation error, $\tilde{D}^{t_j}_{\textnormal{max}}$, for example, ${\tilde{D}^{t_j}_{\textnormal{max}} = \overline{D}^{t_j} - \underline{D}^{t_j}}$. 
  
  To this end, we first determine the bound on the state prediction error from a given past interval and the current state by using the disturbance estimation error bound, $M_d$ from \eqref{pre:est_bounds_m}. Then, we formulate two nonlinear programs to find the set \eqref{pm:max_bounded_set} by enforcing the error bound constraint into the programs. Note that the true delay must satisfy the following state prediction error bound, but since we do not know the true delay here,  we define the state prediction error with the candidate true delay, $D^*$ to be enforced into nonlinear programs.

\subsubsection{State Prediction Error Bound}
Let us consider the difference between the state prediction with the candidate delay $D^*$ and the measured state:
\begin{align}
    e_p(D^*) = ||\bm{x}_p(1,t,D^*) -\bm{x}(t)||, \label{pm:true_error_pre_curr}
\end{align}
where
\begin{align}
    &\bm{x}_p(1,t,D^*) \nonumber\\
    &= \bm{x}(t-\beta)+\beta\int_{0}^{1}{f_0(\bm{x}_p(y,t,D^*),\bm{u}_p(y,t,D^*))}dy, \nonumber
\end{align}
and, ${\bm{u}_p =\bm{u}(t-D^*+\beta(\delta-1))}$, ${ 
 \delta \in [0,1]}$ is a distributed input. For the sake of brevity, we omit the nested functions of $(y,t,D^*)$ and $(y,t,\hat{D})$ in the following. The main idea is that if we formulate the effect of input delay as a disturbance, $e_p$ can be bounded by the disturbance estimation error bound from \eqref{pre:est_bounds_m}. We have the following lemma under the same assumptions of \textit{Lemma~\ref{pre:bound_d_e}}: 
 %can be obtained:
\textit{\begin{lemma}
Let us define the disturbance for system \eqref{eq:sys_unc} as ${\bm{d}(t) = \bm{u}(t-D) - \bm{u}(t-\hat{D})}$, such that $\bm{d}(t), \dot{\bm{d}}(t)$ are uniformly bounded, and consider the disturbance observer is given in \eqref{pm:dob_delay}. Then, the state prediction error is bounded as
\begin{equation}
    e_p(D^*) \leq || B || + \beta\int_{0}^{1}\sigma_{\textnormal{max}}(g(\hat{\bm{x}}_p))M_d dy,
\end{equation}
where $$B(t) \!\triangleq\! \bm{x}(t-\beta)+\beta\int_{0}^{1}{f(\hat{\bm{x}}_p)+g(\hat{\bm{x}}_p)\hat{\bm{u}}_p+g(\hat{\bm{x}}_p)\hat{\bm{d}} dy} -\bm{x}(t)$$ is the state prediction error with $\hat{D}$, and  $\sigma_{\textnormal{max}}(\cdot)$ is the maximum singular value of a given matrix,   $M_d$ is from \eqref{pre:est_bounds_m},  $\hat{\bm{d}}$ is the estimated disturbance, and ${\hat{\bm{u}}_p = \bm{u}(t-\hat{D}+\beta(\delta-1))}$ is a distributed input. \textnormal{The detailed proof is provided in the Appendix.}
\label{pro:true_state_error_bd}
\end{lemma}}

\subsubsection{Update Bound Set on $D$}
Based on the state prediction error bound from \textit{Lemma~\ref{pro:true_state_error_bd}}, we design two nonlinear programs to update the set, $\Xi_{t_j}$ given in \eqref{pm:max_bounded_set}:
\textit{\begin{proposition}
   Consider the input delay estimator \eqref{pre:update_law} with ${D \in \Xi_0}$ and a disturbance observer \eqref{pm:dob_delay} with estimation error bound, \eqref{pre:est_bounds_m}. Then ${D\in  [\underline{D}^{t_j}, \overline{D}^{t_j}] \triangleq \Xi_{t_j} }$ for all $t_j$ and ${\Xi_{t_j} \subseteq \Xi_{t_j-1} \subseteq \Xi_{0}}$, where $\underline{D}^{t_j}$ and $ \overline{D}^{t_j}$ are obtained by solving the optimization problems:
   \begin{align}
        \underline{D}^{t_j} &= \argmin_{D}{\quad D}  \label{pm:nlp1}\\
        \text{s.t. } \quad e_p(D)
        &\leq || B || + \beta\int_{0}^{1}\sigma_{\textnormal{max}}(g(\hat{\bm{x}}_p))M_d dy \nonumber \\ 
        \underline{D}^{t_j-1} &\leq D \leq \overline{D}^{t_j-1},\nonumber 
    \end{align}
    \begin{align}
        \overline{D}^{t_j} &= \argmax_{D}{\quad D}  \label{pm:nlp2} \\
        \text{s.t. } \quad e_p(D) 
        &\leq || B || + \beta\int_{0}^{1}\sigma_{\textnormal{max}}(g(\hat{\bm{x}}_p))M_d dy \nonumber \\ 
        \underline{D}^{t_j-1} &\leq D \leq \overline{D}^{t_j-1}.\nonumber 
    \end{align}
   \label{proposition2}
\end{proposition}}

\begin{proof}
     The proof follows \cite{cohen2022robust} (Lemma 4). We show that ${\Xi_{t_j} \subseteq \Xi_{t_j-1} \subseteq \Xi_{0}}$ first and then ${D \in \Xi_{t_j}}$ in the following. Since the constraint, ${\underline{D}^{t_j-1} \leq D \leq \overline{D}^{t_j-1}}$ ensures that $${\underline{D}^{t_j}, \overline{D}^{t_j} \!\in\! [\underline{D}^{t_j-1}, \overline{D}^{t_j-1}] \!\!\implies\!\![\underline{D}^{t_j}, \overline{D}^{t_j}] \!\subseteq [\underline{D}^{t_j-1}, \overline{D}^{t_j-1}]},$$ then it directly follows that ${\Xi_{t_j} \subseteq \Xi_{t_j-1} \subseteq \Xi_{0}}$ for all ${t_j \in \mathbb{N}}$.
     
    Next, we show that ${D \in \Xi_{t_j}}$, ${\forall t_j \in \mathbb{N}}$. We define two sets:
    \begin{align*}
        &\mathcal{D}_{t_j} \triangleq \{D\in\mathbb{R}_{+} \text{ }| \text{ } ||\bm{x}_p(1,t,D) -\bm{x}(t)|| = 0\}, \\
        &\mathcal{D}^+_{t_j}  \!\triangleq\! \{D \!\in\! \mathbb{R}_{+}\text{ } \!|\! \text{ } e_p(D) \!\leq\!  || B || \!+\! \beta\!\!\int_{0}^{1}\!\!\!\sigma_{\textnormal{max}}(g(\hat{\bm{x}}_p))M_d dy \},
    \end{align*}
    and according to the first inequality constraints in \eqref{pm:nlp1} and \eqref{pm:nlp2}, we have ${\Xi_{t_j} \subset  \mathcal{D}^+_{t_j}}$. %, where
   % \begin{align*}
    %    \mathcal{D}^+_{t_j} = \{D\in\mathbb{R}_{>0}\text{ } | \text{ } e_p \leq  || B || + \beta\int_{0}^{1}\sigma_{\textnormal{max}}(g(\hat{\bm{x}}_p))M_d dy \}.
    %\end{align*}
    Subsequently, it implies that ${\mathcal{D}_{t_j} \subseteq \mathcal{D}^+_{t_j}}$, and from ${D\in \mathcal{D}_{t_j}}$, it follows that ${D\in \mathcal{D}^+_{t_j}}$. With the constraint, ${\underline{D}^{t_j-1} \leq D \leq \overline{D}^{t_j-1}}$ ensures that 
 ${\Xi_{t_j} \subset  \mathcal{D}^+_{t_j} \cap 
  \Xi_{t_j-1}}$, which directly follows that ${D \in \Xi_{t_j}}$ if ${D \in \Xi_{t_j-1}}$. Consequently, since we assume that ${D \in \Xi_{0}}$, it follows that ${D \in \Xi_{t_j-1} \implies  D \in \Xi_{t_j}}$ for all ${t_j\in \mathbb{N}}$, which completes the proof.
\end{proof}

\subsubsection{Update Maximum Error Bound of Delay Estimation}
From the updated set, $\Xi_{t_j}$, we calculate the maximum error bound of the delay estimation as
\begin{equation}
    \tilde{D}^{t_j}_{\textnormal{max}} \triangleq \overline{D}^{t_j} - \underline{D}^{t_j},
    \label{maximum_error_bound_of_delay_est}
\end{equation}
which is used to update the state prediction error bound in \eqref{eqn:prediction_error_bd}, and the error bound is imposed in the safety condition in DaCBFs in the following section.

\subsection{DaCBFs with Minimum Conservatism}
\vspace{-0.3cm}
\textit{\begin{corollary}
    Let $||\bm{u}(t-D)-\bm{u}(t-\hat{D})||\leq \epsilon_{\text{max}}$, and let $\bm{u}(t)$, $\bm{\dot{u}}(t)$ be uniformly bounded from \textit{Lemma~1} for all $t\geq0$. Then the function $h: \mathcal{X}  \rightarrow \mathbb{R} $ is a \textit{Delay adaptive Control Barrier Function} (DaCBF) for \eqref{pf:sys_real} on $\mathcal{S}$ provided that there exists an extended class $\mathcal{K}_{\infty}$ function $\alpha$ such that for all the predicted states, $\bm{\hat{x}}_p \triangleq \bm{x}(t+\hat{D}) \in \mathcal{S}$: 
    \begin{equation}
        \sup_{\bm{u}\in\mathcal{U}}[L_{f}h(\bm{\hat{x}}_p) + L_{g}h(\bm{\hat{x}}_p)\bm{u}(t) -d_e(t)] %\nonumber \\
        \geq -\alpha\big(h(\bm{\hat{x}}_p)\big),
        \label{pre:dacbf_def_new}
    \end{equation}
    where %$\bm{z} = \bm{p}(\bm{x}_p)$ from \eqref{eqn:sensor_measurement}.
    \begin{align}
        d_e(t)=(\mathfrak{L}_{L_fh}+ \mathfrak{L}_{\alpha\circ h})e_{t_j,\text{max}}+\mathfrak{L}_{L_gh}e_{t_j,\text{max}}||\bm{u}(t)|| \label{corrollary_theo1_d}
    \end{align}
    with $e_{t_j,\text{max}}=e_{\text{max}}(t+\hat{D}+\tilde{D}^{t_j}_{\text{max}})+\Delta y_{\text{max}}$. If $h$ is a DaCBF and there exists an input ${\bm{u}\in\mathcal{U}}$ satisfying \eqref{pre:dacbf_def_new}, then the system \eqref{pf:sys_real} is safe w.r.t $\mathcal{S}$ such that ${\forall t \geq D}$.
\end{corollary}}
\begin{proof}
The proof of this corollary follows directly from \cite{YK2024ECC} (Theorem 5). Thus, the detailed proof is omitted for brevity. We remark that an upper bound for ${||\bm{u}(t-D)-\bm{u}(t-\hat{D})||}$ can be constructed with $\tilde{D}_{\textnormal{max}}$ and ${||\bm{\dot{u}}(t)||\leq \dot{u}_{\textnormal{max}}}$ as ${||\bm{u}(t-D)-\bm{u}(t-\hat{D})||}$ ${ = \int_{D}^{\hat{D}} \dot{\bm{u}}(t)dt \leq \dot{u}_{\textnormal{max}}(D-\hat{D}) \leq \dot{u}_{\textnormal{max}}\tilde{D}_{\textnormal{max}}}$.
\end{proof}
\begin{figure*}[t]
    \centering
    \includegraphics[width=1\linewidth]{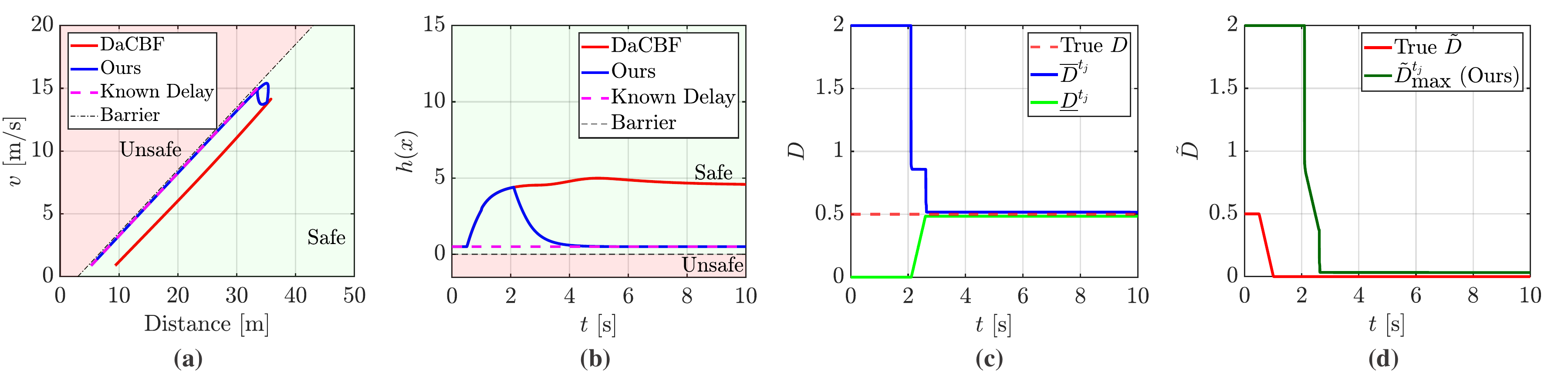}
    \vspace{-0.7cm}
    \caption{The performance of the proposed method and DaCBFs \cite{YK2024ECC} in an automated connected vehicle application under the condition, $D=0.5$. (a) shows the safety regulations of each method. (b) presents barrier function values of each method, indicating the performance of conservatism. (c) and (d) show the results of the two nonlinear programs from \eqref{pm:nlp1} and \eqref{pm:nlp2}, and the plots show that the set, $\Xi_{t_j}$ is gradually reduced, including the true delay $D$. Note that the update algorithm initiates after $2$ seconds because we assume that the initial $\Tilde{D}_{\textnormal{max}}$ is $2$.}
    \vspace{-0.4cm}
    \label{sim:graph}
\end{figure*}
\subsection{Analysis of Conservatism}
In previous work \cite{YK2024ECC}, we used a constant maximum bound of the delay estimation error, ${\tilde{D}^0_{\textnormal{max}}}$, to make the safe controller robust against the state prediction error. Instead of using the constant maximum bound, we update the maximum bound of the estimation error by using \textit{Proposition~\ref{proposition2}} and \eqref{maximum_error_bound_of_delay_est}, and this reduces the conservatism.
\textit{\begin{lemma}
Under the same assumptions of \textit{Proposition}~\ref{proposition2}, let $\tilde{D}^{t_j}_{\textnormal{max}}$ be determined from \eqref{pm:nlp1} and \eqref{pm:nlp2}. Then, 
%the following inequality is obtained 
over sequences, $t_j$, we have: 
\begin{align}
&e_{\text{max}}(t_0+\hat{D}+\tilde{D}^{t_j}_{\textnormal{max}})
\leq e_{\text{max}}(t_0+\hat{D}+\tilde{D}^0_{\textnormal{max}}).
\end{align}
\label{less_conser}
\end{lemma}}
\vspace{-0.35cm}
\begin{proof}
Consider the state prediction maximum errors with $\tilde{D}^{t_j}_{\textnormal{max}}$ from \eqref{pm:nlp1} and \eqref{pm:nlp2} and $\tilde{D}^0_{\textnormal{max}}$ by \cite{YK2024ECC}, respectively:
\begin{align}
&e_{\text{max}}(t_0+\hat{D}+\tilde{D}^{t_j}_{\textnormal{max}})+\Delta y_{\text{max}} \label{pm:error_with_up_max}, \\
&e_{\text{max}}(t_0+\hat{D}+\tilde{D}^0_{\textnormal{max}})+\Delta y_{\text{max}}.  \label{pm:error_with_c_max}
\end{align}
Since the function, $e_{\text{max}}(t)$ from \eqref{eqn:bound_e_dmax} is monotonically increasing, we can show that $\tilde{D}^{t_j}_{\textnormal{max}} \leq \tilde{D}^0_{\textnormal{max}}$ and then we have ${e_{\text{max}}(t_0+\hat{D}+\tilde{D}^{t_j}_{\textnormal{max}})
\leq e_{\text{max}}(t_0+\hat{D}+\tilde{D}^0_{\textnormal{max}})}$. From \textit{Proposition~\ref{proposition2}}, we have ${\Xi_{t_j} \subseteq \Xi_{t_j-1} \subseteq \Xi_{0}}$, which implies that ${\overline{D}^{t_j} - \underline{D}^{t_j} \leq \overline{D}^{t_j-1} - \underline{D}^{t_j-1}}$. Next, from \eqref{maximum_error_bound_of_delay_est}, it inductively follows ${\tilde{D}^{t_j}_{\textnormal{max}} \leq \tilde{D}^{t_j-1}_{\textnormal{max}}}$; thus, it implies: ${\tilde{D}^{t_j}_{\textnormal{max}} \leq \tilde{D}^{0}_{\textnormal{max}}}$ if $\tilde{D}^{0}_{\textnormal{max}}$ holds as an initial condition. Then $e_{\text{max}}$ with the updated maximum bound  \eqref{pm:error_with_up_max} % is less than and equal to \eqref{pm:error_with_c_max}:
satisfies: 
\begin{align}
&e_{\text{max}}(t_0+\hat{D}+\tilde{D}^{t_j}_{\textnormal{max}})
\leq e_{\text{max}}(t_0+\hat{D}+\tilde{D}^0_{\textnormal{max}}),
\end{align}
implying that using \eqref{pm:error_with_up_max} in \eqref{pre:dacbf_def_new} is less conservative than using \eqref{pm:error_with_c_max} proposed by \cite{YK2024ECC}. We remark that the set, $\Xi_{t_j}$, remains constant if (e.g.) $\bm{u}(t)$ is constant.
\end{proof}

\section{Simulation Results}\label{sec:simulation}
In this section, we evaluate the proposed method in an automated connected vehicle application where two trucks are connected via wireless communication and the following truck must maintain a safe distance from the lead vehicle. In this scenario, an unknown input delay can be induced by the communication. Simulations for each method are under the same conditions where the initial input delay estimate is zero, but the true delay, $D$ is $0.5$ seconds. We analyze each method in the context of conservatism based on how far the CBF $h$ is from the boundary of the safe set.

% In this section, we evaluate the proposed method in an automated connected vehicle application where two trucks are connected via vehicle-to-vehicle (V2V) communication and the following truck must maintain a safe distance to the lead vehicle. In this scenario, an unknown input delay can be induced by the communication. To ensure fair comparisons, we consider the same system model, nominal controller, and safety constraint as used in \cite{Tamas2022ISSwithInputDelay}, \cite{YK2024ECC}. We implement DaCBFs and the proposed method under the same conditions where the initial input delay estimate is zero, but the true delay, $D$ is $0.5$ seconds. The delay estimation parameters and $\alpha(\cdot)$ in each method are identical to provide clear comparisons between them. We analyze each method in the context of conservatism based on how far the CBF $h$ is from the boundary of the safe set.

\subsection{Application to an Automated Connected Vehicle Control}
Let us consider the following system dynamics with an unknown, but constant input delay, $D$:
\begin{equation}
\dot{\bm{x}} = \begin{bmatrix}
\dot{\xi} \\
\dot{v} \\
\dot{v}_L
\end{bmatrix} = \underbrace{\begin{bmatrix}
v_L-v \\
0 \\
a_L
\end{bmatrix}}_{f(\bm{x})} +\underbrace{\begin{bmatrix}
0\\
1\\
0
\end{bmatrix}}_{g(\bm{x})}\bm{u}(t-D), \label{sim:system_model}
\end{equation}
where $\xi$ represents the distance between two trucks, $v$ denotes the velocity of the truck behind, and the velocity and acceleration of lead truck are denoted by $v_L, a_L$, respectively. The safety CBF is defined as $h(\bm{x}) = \xi -\xi_{\text{sf}} -Tv$ where $\xi_{\text{sf}}$ denotes a minimum distance for stop, and $T$ represents headway time. We leverage the nominal controller for the following truck proposed in \cite{Tamas2022ISSwithInputDelay}, $\bm{u}_{\text{nom}} = A(V(\xi)-v) + B(W(v_L)-v)$
where $V(\xi) \triangleq \min \{k(\xi - \xi_{\text{st}}), v_{\text{max}}\}$,  $W(v_L) \triangleq \min \{v_L, v_{\text{max}}\}$, and $A$ is an adjustable distance gain, and $B$ denotes a gain for velocity, and $V(\cdot), W(\cdot)$ are the control policies.

The simulation results of each delay adaptive safety controller are shown in Fig.~\ref{sim:graph}. We implement DaCBFs as a baseline without updating the maximum delay estimation error bound, $\Tilde{D}_{\textnormal{max}}$. It is observed that DaCBFs ensure safety but behave conservatively, as shown in Fig.~\ref{sim:graph}(a) since a long distance to the lead truck is maintained. The conservatism is observed in the high positive $h$ function values as shown in Fig.~\ref{sim:graph}(b). In contrast, the proposed method is less conservative than DaCBFs as shown in Fig.~\ref{sim:graph}(a) and (b). This is because the proposed method updates the input delay bound and the maximum delay estimation error bound in an online fashion, as shown in Fig.~\ref{sim:graph}(c) and (d), and consequently, the maximum bound of the state prediction error bound in \eqref{corrollary_theo1_d} is gradually reduced. As a result, the overall conservatism is considerably reduced as indicated in \textit{Lemma~\ref{less_conser}}. In addition, we further evaluate the conservatism of each method in various input delay cases, comparing the average values of $h$ functions in each implementation. 
\begin{table}[h]
\vspace{0.25cm}
\begin{tabular}{c|c|c|c|c|c|c}
\hline
\hline
Input delay  & $0.1s$ & $0.2s$ & $0.3s$ & $0.4s$ & $0.5s$ & Avg.  \\ \hline
DaCBFs \cite{YK2024ECC}     & $4.003$ & $4.195$ & $4.39$ & $4.58$ & $4.77$ & $4.387$ \\ \hline
Ours        & $\bm{0.506}$ & $\bm{0.507}$ & $\bm{0.51}$ & $\bm{0.51}$ & $\bm{0.51}$ & $\bm{0.508}$ \\ \hline\hline
\end{tabular}
\caption{The average values of $h$ functions are shown in each method.}
\vspace{-0.45cm}
\label{table_1}
\end{table}
As indicated in Table.~\ref{table_1}, the proposed method has lower positive values of $h$ functions than the values of DaCBFs across all cases. Furthermore, the average of $h$ functions in all cases is quantitatively lower than DaCBFs when using the proposed method, showing that the conservatism is decreased by $99.78\%$ compared to DaCBFs.
\section{Conclusion}\label{sec:conclusions}
In this paper, we aimed to reduce the conservatism of DaCBFs that results from the maximum state prediction error bound derived with the maximum delay estimation error bound; thus we devised an online adaptive algorithm to update the maximum delay estimation error bound. To this end, we first treated the impacts of the input delay as disturbances and used a disturbance observer with the estimation error bound. The error bound and delay estimation were used to bound the current state prediction error from the previous state. As the input delay was estimated, nonlinear programs updated the maximum delay estimation error bound satisfying the obtained state prediction error bound. Consequently, the updated maximum delay estimation error bound was employed to update the maximum state prediction error bound used in DaCBFs, which led to less conservatism. We verified the proposed method in an automated connected truck application under different settings, showing less conservatism compared to DaCBFs. Potential future directions are to consider time-varying input delay and model uncertainties in systems to achieve safety.

\section{Appendix}
\begin{customproof}[Proof of Lemma~\ref{pro:true_state_error_bd}]
     The state prediction error $e_p$ from \eqref{pm:true_error_pre_curr} can be expressed as 
    \begin{equation}
        e_p(D^*)=||\bm{x}(t-\beta) +\beta\int_{0}^{1}{\dot{\hat{\bm{x}}}_p+g(\hat{\bm{x}}_p)\bm{d}~\!dy} -\bm{x}(t)||, \label{pm:true_error_pre_curr_d}
    \end{equation}
    where ${\dot{\hat{\bm{x}}}_p = f(\hat{\bm{x}}_p)+g(\hat{\bm{x}}_p)\hat{\bm{u}}_p}$, and ${\bm{d} = \bm{u}_p  - \hat{\bm{u}}_p}$. We add and subtract $\beta\int_{0}^{1}g(\hat{\bm{x}}_p)\hat{\bm{d}}dy$ in \eqref{pm:true_error_pre_curr_d} which yields:
    \begin{flalign*}
        e_p(D^*) &= ||\bm{x}(t-\beta)+\beta\int_{0}^{1}f(\hat{\bm{x}}_p)+g(\hat{\bm{x}}_p)\hat{\bm{u}}_p+g(\hat{\bm{x}}_p)\hat{\bm{d}} &\nonumber \\ 
        &\quad \quad \quad \quad  \quad \quad  \quad \quad  \quad +g(\hat{\bm{x}}_p)(\bm{d}-\hat{\bm{d}}) dy -\bm{x}(t)||& \\
        &\leq || B || + ||\beta\int_{0}^{1}g(\hat{\bm{x}}_p)(\bm{d}-\hat{\bm{d}}) dy|| & \nonumber \\
        &\leq || B || + \beta\int_{0}^{1}\sigma_{\textnormal{max}}(g(\hat{\bm{x}}_p))M_d dy, &
    \end{flalign*}
%where $$B \triangleq \bm{x}(t-\beta)+\beta\int_{0}^{1}{f(\hat{\bm{x}}_p)+g(\hat{\bm{x}}_p)\hat{\bm{u}}_p+g(\hat{\bm{x}}_p)\hat{\bm{d}} dy} -\bm{x}(t),$$ and 
%where $M_d$ is from \eqref{pre:est_bounds_m}. Consequently, the state prediction error, $e_p$, satisfies:
%\begin{align*}
%    e_p(D^*) \leq || B || + \beta\int_{0}^{1}\sigma_{\textnormal{max}}(g(\hat{\bm{x}}_p))M_d dy,
%\end{align*}
which is the statement of the lemma.  
\end{customproof}

\section*{Acknowledgement}
The work was supported by Fabrikant Vilhelm Pedersen og Hustrus Legat.
%%%%%%%%%%%%%%%%%%%%%%%%%%%%%%%%%%%%%%%%%%%%%%%%%%%%%%%%%%%%%%%%%%%%%%%%%%%%%%%%
%\section*{APPENDIX}
%\section*{ACKNOWLEDGMENT}
%%%%%%%%%%%%%%%%%%%%%%%%%%%%%%%%%%%%%%%%%%%%%%%%%%%%%%%%%%%%%%%%%%%%%%%%%%%%%%%%

\bibliographystyle{IEEEtran}
\bibliography{bibliography}
\end{document}